\documentclass[apj]{emulateapj}

\shorttitle{Initiation of CMEs}
\shortauthors{Yeates et al.}

\begin{document}


\title{Initiation of Coronal Mass Ejections in a Global Evolution Model}


\author{A. R. Yeates\altaffilmark{1}}
\affil{Harvard-Smithsonian Center for Astrophysics, 60 Garden Street, Cambridge, MA 02138}
\email{ayeates@cfa.harvard.edu}
\altaffiltext{1}{Previously at School of Mathematics and Statistics, University of St. Andrews, St. Andrews, KY16 9SS, UK}

\and
\author{D. H. Mackay}
\affil{School of Mathematics and Statistics, University of St. Andrews, St. Andrews, KY16 9SS, UK}
\email{duncan@mcs.st-and.ac.uk}

\begin{abstract}
Loss of equilibrium of magnetic flux ropes is a leading candidate for the origin of solar coronal mass ejections (CMEs). The aim of this paper is to explore to what extent this mechanism can account for the initiation of CMEs in the global context. A simplified MHD model for the global coronal magnetic field evolution in response to flux emergence and shearing by large-scale surface motions is described and motivated. Using automated algorithms for detecting flux ropes and ejections in the global magnetic model, the effects of key simulation parameters on the formation of flux ropes and the number of ejections are considered, over a 177-day period in 1999. These key parameters include the magnitude and sign of magnetic helicity emerging in active regions, and coronal diffusion. The number of flux ropes found in the simulation at any one time fluctuates between about 28 and 48, sustained by the emergence of new bipolar regions, but with no systematic dependence on the helicity of these regions. However, the emerging helicity does affect the rate of flux rope ejections, which doubles from $0.67$ per day if the bipoles emerge untwisted to $1.28$ per day in the run with greatest emerging twist. The number of ejections in the simulation is also increased by 20\% -- 30\% by choosing the majority sign of emerging bipole helicity in each hemisphere, or by halving the turbulent diffusivity in the corona. For reasonable parameter choices, the model produces approximately 50\% of the observed CME rate. This indicates that the formation and loss of equilibrium of flux ropes may be a key element in explaining a significant fraction of observed CMEs.
\end{abstract}


\section{Introduction}

Coronal mass ejections (CMEs) from the Sun are important drivers of space weather \citep{schwenn2006}. They are believed  fundamental to the long-term evolution of the global solar magnetic field, both by shedding magnetic flux and helicity \citep{low1996,bieber1995,lynch2005}, and by their role in the coronal magnetic field reversal every 11 years \citep[{\it e.g.},][]{zhang2001,owens2007}.

A variety of models have been proposed for the CME initiation process \citep[see the reviews by][]{forbes2000,klimchuk2001,low2001}. Although models differ in their basic magnetic field configuration and in how the eruption is initiated, there is agreement that the driving energy must originate in the magnetic field \citep{forbes2000}. The majority of recent work on CME initiation favours ``storage-and-release'' models \citep{klimchuk2001}, where free magnetic energy is built up gradually in a quasi-static evolution, before sudden release in a highly dynamic eruptive phase. There are essentially two pre-eruption configurations invoked to store the free energy: sheared arcades \citep[{\it e.g.},][]{antiochos1999} or magnetic flux ropes. The latter are twisted structures of helical magnetic fields, anchored at both ends in the photosphere. Note that the degree of twist may be low, and flux ropes with less than one turn are essentially the same as sheared arcades. Indeed, many CME models work with either a flux rope or an arcade \citep{klimchuk2001}; what matters is that sufficient magnetic pressure builds in the structure to overcome the stabilizing magnetic tension of the overlying field. There is, however, a considerable body of evidence that at least a third of CMEs have a flux rope structure, ranging from white-light coronagraph and EUV observations \citep[{\it e.g.},][]{illing1986,dere1999,gibson2006b} to {\it in situ} observations of rotating magnetic field patterns in interplanetary magnetic clouds \citep{klein1982}. Clearly twisted structures are observed in images of erupting prominences at various wavelengths \citep{tandberghanssen1974,plunkett2000,green2007}. It is possible that flux ropes observed in CMEs may be formed \emph{during} eruption \citep{gosling1995} rather than existing previously in a quiescent state. However, the existence of coronal cavities before lift-off is strong evidence for a pre-existing flux rope in many cases \citep{gibson2006b}, because the flux rope can explain both the lower density of the cavity (due to strong magnetic pressure), and the support of filament mass inside the cavity \citep{gibson2006}.

Previous models for CME initiation have considered only a single event in a simplified magnetic configuration, in order to study the basic physical process. In this paper, however, we begin to address the fundamental question of where and when CMEs occur in the global context. A quasi-static model is used to study the build up of axial flux in magnetic flux ropes, using observed photospheric flux distributions as input to drive simulations of the global coronal magnetic field. In essence, this model is an extension of the two-dimensional (2D) catastrophe models considered by \citet{vantend1978} and many authors since \citep[see the review by][]{lin2003}. These 2D catastrophe models describe the equilibrium curve of a flux rope as some control parameter, typically describing the background magnetic field, is varied. Eventually, a nose point in the equilibrium curve may be reached beyond which no equilibrium is possible. In the present 3D global model, many flux ropes form at different locations on the Sun, and may or may not lose equilibrium as they evolve. The location of each flux rope and evolution of its background field are determined self-consistently based on the observed photospheric magnetic field, rather than being prescribed as in the 2D models. Therefore, in this paper, the evolution of the coronal field leading to the formation and ejection of flux ropes is constrained by observations.

The global model described in this paper differs from other existing models of the global magnetic field in the solar corona. The majority of such models have used potential-field extrapolations with the radial magnetic field in the solar photosphere as a lower boundary condition, and an upper source surface at $r=2.5 R_\odot$ \citep{altschuler1969,schatten1969}. To study the time evolution of the corona with potential-field models, a sequence of independent extrapolations has been used \citep[{\it e.g.},][]{wang2002,mackay2002,schrijver2003}. Unfortunately, the potential field suffers from the fundamental limitation of having the minimum magnetic energy for the given boundary conditions, thus containing no free magnetic energy to drive CMEs. More general extrapolation methods are under development for the global corona, including nonlinear force-free fields based on full-disk vector magnetograms \citep{wiegelmann2007} and full-MHD solutions \citep[{\it e.g.},][]{mikic1999,riley2006}. However, such models are too computationally intensive to simulate the long-term evolution (over periods of months or longer), so we have developed a new technique using a coupled surface flux transport and magneto-frictional model \citep[][hereafter ``Paper I'' and ``Paper II'']{yeates2007,yeates2008a}. This allows us to insert active regions with non-zero magnetic helicity into the global corona, and to follow the evolution of twisted and sheared non-potential magnetic fields through a sequence of equilibria as they are driven by photospheric motions. Like the 2D catastrophe models, a quasi-static approximation of the evolution is used, so that the model can follow the build up of flux ropes and their loss of equilibrium, but not the subsequent dynamics of the eruption, which would require full time-dependent MHD simulations. Note that coronal flux ropes form not only within active regions, but also between regions, consistent with the observations of solar filaments \citep{tang1987,mackay2008}. The research presented here, in considering the formation and ejection of flux ropes, has been motivated by the success of the model in reproducing the hemispheric chirality pattern of sheared magnetic fields in filaments (Paper II) as non-potential fields of solar filaments are known to be related to CMEs.

The aim of this paper is to describe the model and its physical motivation, and to demonstrate how the formation and evolution of magnetic flux ropes in the model depends on certain parameters, in particular the magnetic helicity of emerging regions. We physically test whether flux cancellation, as simulated in this model, can produce enough flux rope ejections to match the observed CME rate, or whether other physical mechanisms are required. A detailed comparison of our simulation with the observed locations of CMEs will be the subject of a future paper. The organization of this paper is as follows. The model and its physical assumptions are described in Section \ref{sec:model}. In Section \ref{sec:theory}, we describe the basic process of flux rope formation and ejection that occurs many times in the global model. Automated techniques for detecting flux ropes and their ejection in the global simulations are described in Section \ref{sec:auto}, then applied in Section \ref{sec:results} to consider the effect of simulation parameters on the flux ropes formed and on the rate of ejections. In Section \ref{sec:conclusions} we summarize our conclusions, and suggest future extensions of the work.

\section{Coronal Evolution Model} \label{sec:model}

A full treatment of the dynamical evolution of the global solar corona over weeks to months is computationally prohibitive. Instead, we use the coupled flux transport and magneto-frictional model of \citet{vanballegooijen2000}, which describes the evolution of the large-scale mean magnetic field under certain assumptions. The basic physics of the model has been studied in a series of papers with the aim of understanding the chirality of observed solar filaments. These papers include detailed parameter studies of a pair of interacting bipolar magnetic regions \citep{mackay2001,mackay2005,mackay2006a,mackay2006b}, in addition to simulations driven by observed photospheric magnetic fields, both local \citep{mackay2000}, and global \citep[Paper I, Paper II, ][]{yeates2008b}, as in this paper. In this section we summarize the basic assumptions of this model, before describing our particular numerical implementation. We describe the observational input required for the simulations to accurately reproduce the photospheric and coronal magnetic fields.

\subsection{Basic Physical Assumptions} \label{sec:assumptions}

To model the evolution of the large-scale magnetic field in the solar corona, we make the following simplifying assumptions:
\begin{enumerate}
\item The magnetic field in the solar corona may be separated into a large-scale, mean, component on lengthscales $\sim 50\,\textrm{Mm}$, and a fluctuating component on lengthscales $\sim 1\,\textrm{Mm}$.
\item On the timescale of large-scale photospheric motions, the large-scale coronal magnetic field is in equilibrium, except perhaps during flares or CME eruptions.
\item Above the photosphere, and in the low corona up to $2.5R_\odot$, magnetic forces dominate over plasma forces ({\it i.e.}, it is a low-$\beta$ plasma). In the photosphere magnetic flux tubes are passively advected.
\item The large-scale magnetic field in the photosphere is predominantly radial, and its horizontal lengthscale is larger than that of the supergranular convection.
\item For purposes of the large-scale magnetic field evolution, emerging active regions may be represented by twisted magnetic bipoles, with an idealized form.
\end{enumerate} 
Assumption 1 allows the application of mean field theory to the corona \citep{seehafer1994,vanballegooijen2000}. In that case, only the mean magnetic field is solved for, and the effect of significant small-scale structure---such as braiding and current sheets produced by interaction with convective flows in the photosphere \citep{parker1972}---is included through a turbulent electromotive force (e.m.f.). The mean-field induction equation takes the form
\begin{equation}
\frac{\partial\mathbf{A_0}}{\partial t} = \mathbf{v}_0\times\mathbf{B}_0 + \mathcal{E}_0,
\label{eqn:induction}
\end{equation}
where $\mathbf{A}_0(\mathbf{r},t)$ is the vector potential for the mean magnetic field $\mathbf{B}_0 = \nabla\times\mathbf{A}_0$, $\mathbf{v}_0(\mathbf{r},t)$ is the mean plasma velocity, and $\mathcal{E}_0(\mathbf{r},t)$ is the turbulent e.m.f..

Assumption 2 is valid because magnetic disturbances from the photosphere propagate up into the corona at Alfv\'{e}n speeds of the order $1000\,\textrm{km}\,\textrm{s}^{-1}$ or greater \citep{regnier2008}, well above the maximum flow speed of surface rotation (of the order $0.2\,\textrm{km}\,\textrm{s}^{-1}$). The evolution of the coronal magnetic field may then be understood as a continual distortion of existing equilibrium by footpoint motions, with subsequent fast relaxation to a neighbouring equilibrium \citep{seehafer1994}. By assumption 3, the form of this equilibrium is well approximated by a force-free field $\mathbf{j}_0\times\mathbf{B}_0=0$, where $\mathbf{j}_0=\nabla\times\mathbf{B}_0/\mu_0$ is the mean current density.

In contrast to the corona, the photospheric field is assumed to be passively advected by plasma flows. Assumption 4 then justifies use of the standard surface flux transport model for the evolution of the radial component $B_{0r}$ on the solar surface \citep[][and references therein]{sheeley2005}. That the large-scale photospheric field should be approximately radial is supported both by vector magnetic measurements \citep{martinezpillet1997}, and by theoretical considerations, due to the concentration of convection zone magnetic flux into isolated kilogauss flux tubes \citep{vanballegooijen2007}. In the surface flux transport model, $B_{0r}$ is advected by the large-scale motions of differential rotation and meridional circulation, and the influence of supergranular convection on this large-scale field is modelled by a surface diffusion term \citep{leighton1964}. In the model described in this paper, surface flux transport, coupled with the emergence of new active regions, acts as the lower boundary condition to drive the coronal field evolution.

Assumption 5 is the most uncertain, due to the idealized nature of the magnetic bipoles. Previous studies have shown that such idealized bipoles are adequate for reproducing the distribution of large-scale radial field $B_{0r}$ on the solar surface (Paper I). The main uncertainty is the inclusion of twist. While our implementation in this paper is simplistic, it is better than not including any twist. Such twist has been shown to be required both by vector magnetic field measurements \citep{demoulin2007} and by our previous simulations (Paper II). Part of the aim of the present work is to determine whether such a simplified model can yield useful information about the global magnetic field, or whether more detailed modelling of individual regions is necessary to capture even the simplest large-scale phenomena.

\subsection{Numerical Implementation}

As in Paper II, we solve Equation (\ref{eqn:induction}) in a domain extending from $0^\circ$ to $360^\circ$ in longitude, $-80^\circ$ to $80^\circ$ in latitude, and $R_\odot$ to $2.5R_\odot$ in radius. We assume the turbulent e.m.f. to take the form
\begin{equation}
\mathcal{E}_0 = -\eta\mathbf{j}_0,
\end{equation}
with the turbulent diffusivity
\begin{equation}
\eta = \eta_0\left( 1 + 0.2\frac{|\mathbf{j}_0|}{|\mathbf{B}_0|}\right)
\label{eqn:eta}
\end{equation} 
as in \citet{mackay2006a}. The first term is a uniform background value and the second term is an enhancement in regions of strong current density. This enhancement was introduced because earlier simulations \citep{mackay2001} produced highly twisted, structures which are typically not observed \citep{bobra2008,su2009}.

To follow the evolution through a sequence of nonlinear force-free states in response to flux emergence and surface shearing (Section \ref{sec:assumptions}), the MHD momentum equation is approximated by the magneto-frictional method \citep{yang1986}, setting
\begin{equation}
\mathbf{v}_0 = \frac{1}{\nu}\frac{\mathbf{j}_0\times\mathbf{B}_0}{B^2} + v_\textrm{out}(r)\hat{\mathbf{r}}.
\label{eqn:v}
\end{equation}
The second term in Equation (\ref{eqn:v}) is a radial outflow imposed near the upper boundary, where it simulates the effect of the solar wind in opening up field lines in the radial direction \citep{mackay2006a}. It has a peak value $v_0=100\,\textrm{km}\,\textrm{s}^{-1}$ and falls off exponentially away from the upper boundary, so that only structures near this boundary are affected.

We assume closed boundaries in the latitudinal direction and an open boundary at the upper source surface ($r=2.5R_\odot$), where the radial outflow ensures that the magnetic field is radial. At the lower boundary ($r=R_\odot$) the radial magnetic field $B_{0r}$ is evolved according to the surface flux transport model with flux emergence, as described in Paper I.

In this paper, we simulate the solar corona over 177 days in 1999, the same period as Paper II. The initial condition is a potential field extrapolation for 1999 April 16 (day of year 106) taken from a Kitt Peak synoptic magnetogram corrected for differential rotation (Paper I). The coronal magnetic field is then evolved continuously using Equations (\ref{eqn:induction}) and (\ref{eqn:v}), with surface flux transport on the lower boundary and flux emergence, until 1999 October 10 (day of year 283). During this period, 119 new active regions are inserted in the form of 3D twisted magnetic bipoles, to be described in Section \ref{sec:obs}. The model then follows the quasi-static build-up of magnetic energy in the solar corona.

\subsection{Observational Input} \label{sec:obs}

\begin{figure*}
\includegraphics[width=\textwidth]{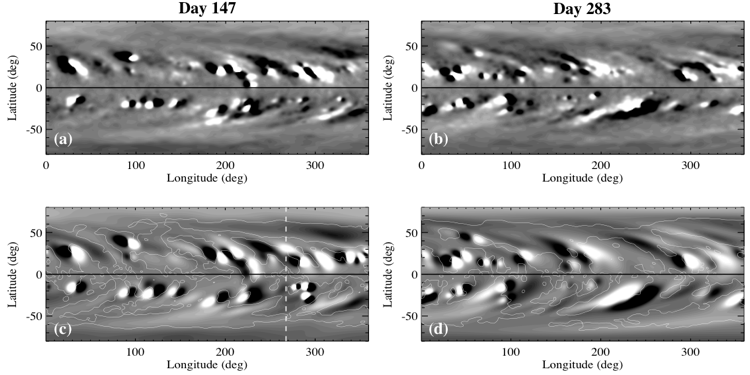}
\caption{Comparison between observed synoptic magnetograms (panels a, b) and simulated radial magnetic field $B_{0r}$ on the photosphere (panels c, d), on day 147 and day 283. White shading indicates positive flux and black shading negative, with a saturation level of $\pm 25\,\textrm{G}$. The observed magnetograms (a, b) are corrected for differential rotation and smoothed for easier comparison. On the simulated magnetograms (c, d), the zero contours from the {\it observed} magnetograms (a, b) are overlayed in white. To the right of the white dashed line, new regions from the next rotation will appear in (c) but not in (a), because they are inserted 7 days before observation at central meridian.}\label{fig:surf}
\end{figure*}

The emerging magnetic bipoles, which constitute the primary observational input to our simulations, are based on active regions observed in synoptic normal-component magnetograms from the National Solar Observatory (NSO) at Kitt Peak. For each of 119 bipoles emerged during this 177-day period, we recorded the day of central meridian passage, latitude, size, tilt angle, and magnetic flux from the Kitt Peak data (Paper I). Rather than using the actual magnetic field distribution of each region, we insert idealized magnetic bipoles---of the mathematical form given in Paper II---into the simulation, with parameters chosen to match the above observed properties. The advantage of this approach is that the bipoles may readily be inserted in 3D with photospheric and coronal components, and non-zero magnetic helicity. To model each observed active region more accurately would require the availability of vector magnetograms, and an extrapolation to model the unknown coronal magnetic field. Methods are under development to extrapolate nonlinear force-free fields from photospheric vector magnetograms, but there are still serious problems with this procedure \citep[see][]{derosa2009}, apart from the major computational cost for a large number of regions. Our idealized bipoles cannot reproduce the detailed structure and dynamics within individual active regions. However, the model has been shown to reproduce both the long-term global dispersal of active regions across the solar surface (Paper I) and the statistical hemispheric pattern of filament chirality (Paper II). The reproduction of the observed surface magnetic field is illustrated in Figure \ref{fig:surf}, which shows the observed and simulated $B_{0r}(R_\odot,\theta,\phi)$ on day 147 (after 41 days' evolution) and on day 283 (at the end of the simulation). The accuracy of the simulated field may be seen from Figures \ref{fig:surf}(c) and \ref{fig:surf}(d), where the zero contours from the observed magnetograms are overlayed on the simulated flux distribution. We now consider two aspects of the idealized bipoles which are poorly determined from observations.

\subsubsection{Bipole Insertion}

Because we only observe one side of the Sun, the exact date of emergence of most active regions is unknown, as are the details of the flux emergence process. For simplicity, we insert each region instantaneously, on an arbitrary  day which we choose to be 7 days before central-meridian passage of the region. As detailed in Paper I, the properties of the bipole are ``evolved'' back in time so as to reproduce the observed properties when passing central meridian. Rather than simply adding the bipole vector potential to any pre-existing magnetic field, we ``sweep'' away pre-existing field from the insertion region (Fig. 3 of Paper II). This creates a more realistic end-state for the emerged region as an independent flux system, and prevents the formation of disconnected flux in the corona. It also crudely models the distortion of pre-existing coronal field by a newly-emerging flux tube, as found in MHD simulations of flux emergence \citep[{\it e.g.},][]{yokoyama1996,krall1998}. Of course, the sweeping procedure has an impact on the structure of sheared magnetic fields around the bipole, and thus on some of the magnetic flux ropes considered in this paper. However, in a detailed study of the origin of sheared fields at 109 filament locations in the simulation \citep[][hereafter ``Paper III'']{yeates2009}, the sweeping was found to be important in only 19 cases.

\subsubsection{Bipole Twist}

\begin{deluxetable*}{crrrr}
\tablewidth{0pt}
\tablecaption{Summary of parameters in different simulation runs.}
\tablehead{
\colhead{Run} & \colhead{$\beta$ in N. hemisphere} & \colhead{$\beta$ in S. hemisphere} & \colhead{$\eta_0$ ($\textrm{km}^2\,\textrm{s}^{-1}$)} & \colhead{$v_0$ ($\textrm{km}\,\textrm{s}^{-1}$)}}
\startdata
AN & \multicolumn{2}{c}{No emerging regions} & 45 & 100\\
Am6 & 0.6 & -0.6 &  45 & 100\\
Am4 & 0.4 & -0.4 &  45 & 100\\
Am2 & 0.2 & -0.2 &  45 & 100\\
A0 & 0 & 0 &  45 & 100\\
A2 & -0.2 & 0.2 &  45 & 100\\
A4 & -0.4 & 0.4 &  45 & 100\\
A6 & -0.6 & 0.6 &  45 & 100\\
D4 & -0.4 & 0.4 & 22.5 & 100\\
V4 & -0.4 & 0.4 & 45 & 50\\
\enddata
\label{tab:runs}
\end{deluxetable*}

\begin{figure}
\includegraphics[width=\columnwidth]{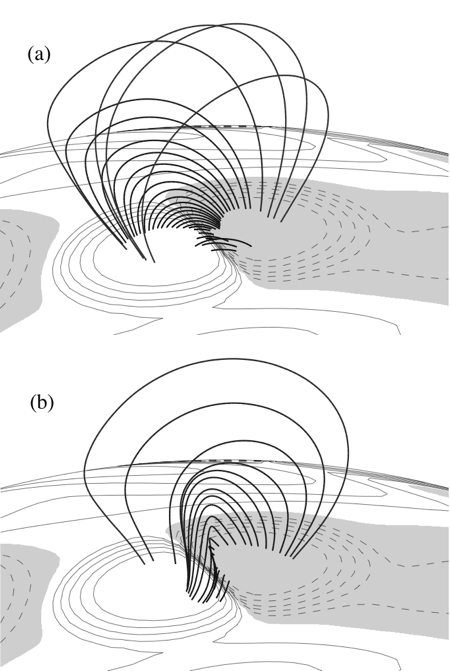}
\caption{Effect of the twist parameter $\beta$ on the shape of a 3D magnetic bipole with (a) $\beta=0.2$ and (b) $\beta=-0.6$. Grey shading and thin contours show the radial magnetic field $B_{0r}$ on the photosphere (white/solid contours for positive, grey/dashed contours for negative). Thick lines show selected coronal magnetic field lines.}\label{fig:bipoles}
\end{figure}

The 3D magnetic bipoles are given a non-zero magnetic helicity through a parameter $\beta$ (defined in Equations 6 and 8 of Paper II), which describes the initial twist of the field lines in the corona, and does not affect the bipole's radial magnetic footprint on the solar surface. Figure \ref{fig:bipoles} shows the structure of a bipole with two different magnitudes and signs of $\beta$, taken from two simulation runs in this paper. The effect of $\beta$ on the flux ropes forming in a simple two-bipole configuration has previously been considered by \citet{mackay2006a}. Here coronal flux ropes formed not only within the two bipoles at sites of flux emergence, but also between the bipoles where flux was cancelling. The magnitude of $\beta$ affected the formation rate of magnetic flux ropes, and hence the time until loss of equilibrium, but did not affect whether or not a flux rope formed. \citet{mackay2005} showed that the sign of $\beta$ also affects the chirality type (direction of shear) in this simple configuration. Subsequent studies of the chirality generated in global simulations have found that both emerging helicity and shearing by surface motions can be important at different locations on the Sun (Paper III). For the 109 filament locations examined in that study, only 32\% changed chirality when the sign of $\beta$ in nearby regions was reversed.

Ideally, the magnitude of helicity in each bipole would be based on vector magnetograph data for the real active region in 3D space, but such measurements are unavailable in the corona. Although a number of authors have attempted to estimate helicity in active regions using vector magnetic field measurements in the photosphere \citep[see][]{demoulin2007}, the horizontal magnetic field components have high uncertainties---including a $180^\circ$ degree ambiguity \citep{metcalf2006}---which are exacerbated by taking derivatives to compute the vertical current density. Thus we are unable to determine observationally the best value of $\beta$ to model each active region. We therefore assume that all bipoles in each hemisphere have the same value of $\beta$, and run simulations with different values to determine the effect this has on the behavior of the model. Note that taking the same value of $\beta$ for all regions in each hemisphere does not lead to uniform magnetic helicity in space as the bipoles have different sizes and field strengths. This is well illustrated by \citet{yeates2008b} in plots of the ``current helicity'' $\alpha=\nabla\times\mathbf{B}_0\cdot\mathbf{B}_0/B_0^2$, a measure of the local twist of field lines in a force-free field. One seemingly robust feature of magnetic helicity in the solar corona is the hemispheric pattern, whereby the majority of active regions in the Northern hemisphere have negative helicity, and in the Southern hemisphere, positive helicity. This is suggested not only by the photospheric vector magnetic measurements \citep{pevtsov1995,abramenko1996,hagino2004}, but also by proxy observations such as H$\alpha$ images of active region structure \citep{hale1927,balasubramaniam2004}, X-ray sigmoids \citep{canfield2007}, and in situ heliospheric measurements \citep{smith1993}. In accordance with this pattern we choose opposite signs of $\beta$ in each hemisphere.

To determine the effect of the $\beta$ parameter on the magnetic flux ropes forming in the global simulation, we consider in this paper a number of simulation runs, as summarized in Table \ref{tab:runs}. These include runs where the emerging bipole twist takes the observed majority sign in each hemisphere (A2, A4, and A6), runs where it takes the opposite sign (Am2, Am4, and Am6), and a run where the bipoles emerge untwisted (A0). For comparison, we also include a run with no emerging bipoles (AN), a run with lower coronal diffusivity (D4), and a run with a lower outflow speed at the upper boundary (V4).

\section{Flux Rope Formation and Evolution} \label{sec:theory}

\begin{figure*}
\includegraphics[width=\textwidth]{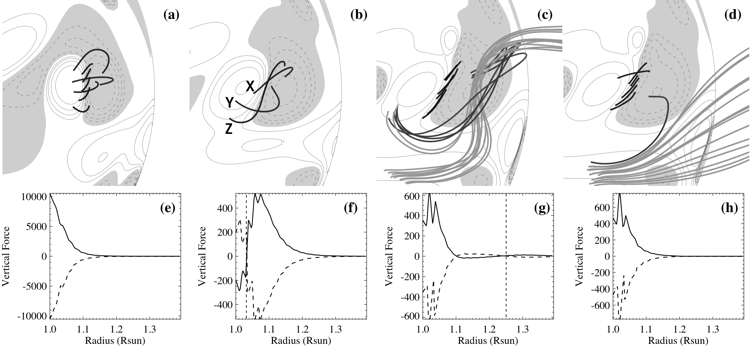}
\caption{Example flux rope forming in a bipolar region in simulation run A2. Top row shows the magnetic field on days (a) 191, (b) 213, (c) 224, and (d) 226. Bottom row (panels e to h) shows vertical components of the magnetic pressure force (solid lines) and magnetic tension force (dashed lines) on the same days, as a function of radius above the center of the bipolar region (in arbitrary units). In panels (a) to (d) grey shading and thin contours show the radial magnetic field $B_{0r}$ on the photosphere, and thick lines are selected coronal field lines (lighter grey denotes open field lines). In panels (f) and (g) the dashed vertical line indicates the height of the flux rope axis.}\label{fig:example}
\end{figure*}

In this section we outline the basic physical processes of magnetic flux rope formation and ejection that are captured in our model, illustrated with an individual flux rope from one of the global simulations. Overall statistics of flux ropes in the global simulations are presented in Section \ref{sec:results}.

Under assumptions 1 to 4 in Section \ref{sec:assumptions}, a robust feature of the coronal evolution is the quasi-static accumulation above photospheric polarity inversion lines (PILs) of axial magnetic field ({\it i.e.}, the component parallel to the PIL). This has been demonstrated in numerical simulations \citep{vanballegooijen1989,vanballegooijen1998,linker2003}. Field lines naturally converge towards PILs because of surface diffusion of their photospheric footpoints. Although there is cancellation of opposite polarity magnetic flux at the PIL \citep{martin1985}, any axial component of magnetic flux must remain in the corona, because the photosphere presents a barrier to the submergence of horizontal field \citep{vanballegooijen2007}. The origin of the axial component may be simple shearing by differential rotation, but it may also arise from the morphology of emerging active regions, particularly if they contain non-zero helicity. Paper III showed that the model used in the present paper is able, over time, to form the hemispheric pattern of axial magnetic fields in filaments \citep{martin1994}, precisely because it allows for these different sources of axial field. More generally, the formation of localized flux ropes in the corona is suggested to be a natural result of the approximate conservation of total magnetic helicity during relaxation to a lower energy state \citep{gibson2006}.

Figure \ref{fig:example} illustrates the formation and ejection of a flux rope from one of our global simulations (run A2). Here, the flux rope forms above the internal PIL of a bipolar region. Panels (a) and (b) show the formation of a flux rope as the footpoints of sheared loops cancel at the PIL. The basic mechanism for formation of a twisted flux rope is described by \citet{vanballegooijen1989}. Diffusivity in the corona allows two loops (such as those labelled X and Y in Figure \ref{fig:example}b) to reconnect, leading to the formation of longer, twisted, axial field lines such as that labelled Z (formed by an earlier cancellation). In this example, an axial magnetic component was present initially because the bipole was inserted with non-zero twist, although it has been enhanced by differential rotation over 22 days.

In an unbounded force-free field outside a sphere (such as the solar photosphere $r=R_\odot$), \citet{zhang2006} suggest that there is an upper limit to the helicity that may accumulate. This arises from the virial theorem for such a field,
\begin{equation}
\int_{r>R_\odot}\frac{B^2}{2\mu_0}dV = \frac{1}{2\mu_0}\int_{r=R_\odot}\left(B_r^2 - B_\theta^2-B_\phi^2\right)\,\textrm{d}\Omega,
\end{equation}
which places an upper limit on the magnitude of the horizontal magnetic field components $B_\theta$, $B_\phi$ on the surface $r=R_\odot$, and thus on the helicity. \citet{zhang2006} verify this conjecture for a particular sequence of axisymmetric solutions. It suggests that a flux rope cannot remain in force-free equilibrium once its axial field becomes too strong. Indeed, the sudden loss of equilibrium as a flux rope gradually increases in size is a robust feature demonstrated both in simple analytical ``catastrophe models'' \citep[see the review by][]{lin2003}, and in full MHD simulations \citep{amari2000,linker2003}. Such sudden losses of equilibrium occur in our model when too much axial flux accumulates in flux ropes over the quasi-static evolution. A key aim of this paper is to determine what proportion of CMEs could be accounted for by this quasi-static model, and what proportion require some other initiation mechanism (Section \ref{sec:conclusions}).

Loss of equilibrium of a flux rope as the axial field becomes too strong has previously been demonstrated in a two-bipole configuration using the present magneto-frictional model \citep{mackay2006a}. It was found that, after equilibrium is lost, the flux rope is driven upward by reconnection underneath, and expelled through the upper boundary of the computational domain. Due to its simplified nature, the model does not follow the true dynamics upon eruption; this would require full MHD simulations that include the inertia of the coronal plasma. However, we believe that the model describes the build-up to eruption accurately. The ejection removes strongly-sheared field lines and magnetic helicity through the upper boundary, leaving only a small residual helicity at low heights. After the ejection, a new flux rope begins to form by flux cancellation at the same location, and this cycle continues. Note that the eruption is localized and regions away from the site of the eruption are unaffected and remain in equilibrium. Figure \ref{fig:example}(c) shows our example flux rope just after equilibrium has been lost and the axis is accelerating upward. Many of the flux rope field lines have already been opened (shaded in lighter grey). Figure \ref{fig:example}(d) shows the configuration after the flux rope has been ejected from the computational domain---note the remaining sheared (but shorter) field lines low in the corona. \citet{mackay2006a} found that the frequency of lift-offs in their simple configuration depended both on the helicity of the bipolar regions and on the strength of the overlying field. Thus in our global simulations we expect the behavior of flux ropes to vary both due to the value of bipole twist $\beta$, and due to the local background magnetic field at each location. We now consider the global simulations.

\section{Automated Detection Methods} \label{sec:auto}

In this section, we describe automated techniques to detect flux ropes (Section \ref{sec:autoropes}) and ejections (Section \ref{sec:autoeject}) in sequences of daily magnetic field snapshots from the global simulations. Global statistics derived with these techniques are then presented in Section \ref{sec:results}.

\subsection{Detection of Flux Ropes} \label{sec:autoropes}

\begin{figure}
\includegraphics[width=\columnwidth]{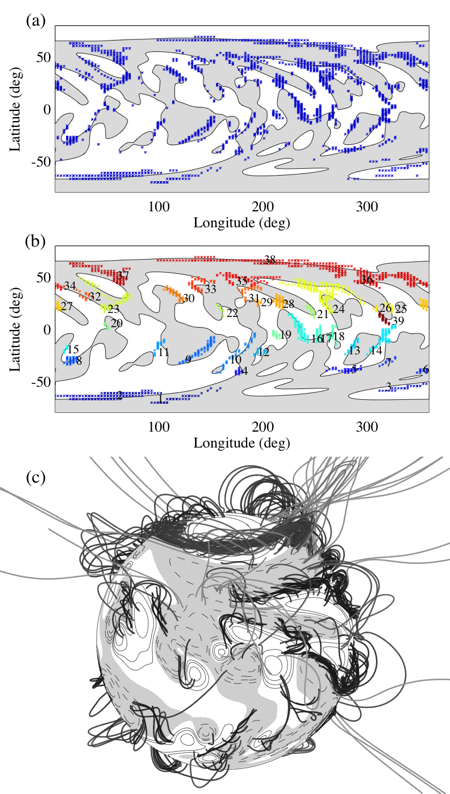}
\caption{Automated detection of magnetic flux ropes in the simulated magnetic field, for day 202 of run A2. The first two panels show (a) selected points and (b) clusters, projected on the normal magnetic field $B_{0r}$ in the photosphere (white shading for positive, grey for negative). Panel (c) shows coronal magnetic field lines traced from the selected flux rope points (with lighter grey denoting open field lines), viewed from $180^\circ$ longitude and $20^\circ$ latitude.}\label{fig:detection}
\end{figure}

The basic idea is to search for the distinctive Lorentz force structure of a flux rope. This is illustrated in Figures \ref{fig:example}(e) to (h) for the example given in Section \ref{sec:theory}. The Lorentz force may be decomposed as
\begin{equation}
\mathbf{j}_0\times\mathbf{B}_0 = \frac{1}{\mu_0}\left(\mathbf{B}_0\cdot\nabla\right)\mathbf{B}_0 - \nabla\left(\frac{B_0^2}{2\mu_0}\right),
\label{eqn:jxb}
\end{equation}
where the first term on the right-hand side is the magnetic tension $\mathbf{T}$ and the second is the magnetic pressure gradient $\mathbf{P}$. Figures \ref{fig:example}(e) to (h) show the vertical components of each of these terms $T_z$ (dashed line) and $P_z$ (solid line), as a function of height above the centre of the PIL. Notice that the two terms are approximately equal and opposite, consistent with a force-free equilibrium with vanishing Lorentz force. The distinctive flux rope structure consists of a sign reversal in both $P_z$ and $T_z$ at the height of the flux rope axis, denoted by the vertical dashed line in Figures \ref{fig:example}(f) and (g). In a flux rope, the helical structure is such that the tension force acts inward toward the rope axis, but the magnetic pressure force acts outward \citep[][see their Fig. 13]{mackay2006a}.

The automated procedure for detecting flux ropes then consists of two stages:
\begin{enumerate}
\item \emph{Point testing:} Test individual points on the numerical grid for the characteristic flux rope structure of sign changes in the vertical magnetic pressure and tension forces.
\item \emph{Clustering:} Use a hierarchical clustering algorithm \citep{johnson1967} to associate neighbouring points which form part of the same flux rope structure.
\end{enumerate}

First consider stage 1. At each point $(r_i,\theta_i,\phi_i)$ in the computational grid the vertical components $P_z$ and $T_z$ of the magnetic pressure and tension forces are computed from the magnetic field $\mathbf{B}_0$. We then require that
\begin {eqnarray}
P_z(r_{i-1},\theta_i,\phi_i) &<& 0,\\
P_z(r_{i+1},\theta_i,\phi_i) &>& 0,\\
T_z(r_{i-1},\theta_i,\phi_i) &>& 0,\\
T_z(r_{i+1},\theta_i,\phi_i) &<& 0
\end{eqnarray}
for the point $(r_i,\theta_i,\phi_i)$ to be selected. In practice, we do not need to test all points, but rather test points only up to height $r=1.44R_\odot$, and use a coarser ``testing grid'' of $(21, 93, 120)$ points in the $(r,\theta,\phi)$ directions. In the $\phi$ direction at the equator this corresponds to one third of the computational grid resolution. The testing grid is chosen such that each point represents an equal 3D volume, by taking equal steps in $\cos\theta$, $\phi$, and $r^3$. Using a coarser grid reduces computational effort and does not affect the results since we are only interested in well-resolved flux rope structures. In order to concentrate the sample on twisted flux ropes, we have implemented the fifth condition of a sufficiently strong parallel current at each point tested, requiring that
\begin{equation}
\frac{\left|\mathbf{j}_0\cdot\mathbf{B}_0\right|}{B_0^2} > \alpha^*,
\end{equation} 
with the threshold $\alpha^* = 0.7\times 10^{-8}\,\textrm{m}^{-1}$ determined by experiment. By way of example, Figure \ref{fig:detection}(a) shows the flux rope points identified by this first stage on day 202 of run A2, mid-way through the simulation. The points are projected on a plot of $B_{0r}$ on the photosphere, showing the PIL dividing regions of positive (white) and negative (grey) magnetic polarity. The identified points tend to lie above PILs, as expected from our basic theory of flux rope formation (Section \ref{sec:theory}). Furthermore, the points are mostly grouped into larger structures. Automated detection of these groupings forms stage 2 of the procedure.

The basic clustering idea is simple. Starting with each point as an individual entity, the two closest points are grouped together, and the process repeated until the shortest inter-group distance (in 3D space) is above some threshold. For this threshold we choose $5\Delta R_\odot$, where $\Delta$ is the heliographic angle in radians of a computational grid cell at the equator. After running this clustering algorithm, any groups with fewer than 8 points on the testing grid are removed. Again this value was determined by visual inspection of the selected structures. The results after clustering for day 202 of run A2 are shown in Figure \ref{fig:detection}(b). Each individual group of points, or ``flux rope'', is identified by color and a number. The actual magnetic field structures corresponding to these flux ropes are illustrated in Figure \ref{fig:detection}(c), which shows 3D magnetic field lines traced from the selected flux rope points in each structure.

\subsection{Detection of Ejections} \label{sec:autoeject}

\begin{deluxetable*}{rccr}
\tabletypesize{\footnotesize}
\tablewidth{0pt}
\tablecaption{Flux rope ejections in run A2 between day 202 and day 214.}
\tablehead{
\colhead{Actual Days} & \colhead{Latitude \& Longitude (deg.)\tablenotemark{a}} & \colhead{Detected Days\tablenotemark{b}} & \colhead{Clustered Points\tablenotemark{c}}}
\startdata
198---203 & 45, 185 & 195---202 (197)& 60\\
202		& 15, 245 & 195---202 (197)& 135\\
202---203 & 10, 315 & 201---203 (202)& 46\\
203---204 & 5, 230 & 200---205 (205)& 28\\
204---206 & 65, 200 & 202---204 (204)& 46\\
205		& -40, 260 & not detected &---\\
205---206 & -20, 195 & 205		& 19\\
206		& 60, 175 & 206---207 (206) & 16\\
206---207 & -35, 270 & 206---207 (206)& 10\\
207---208 & -20, 175 & 207---209 (208)& 10\\
208            & 25, 55   & 208		& 7\\
208---212 & -60, 5	& not detected &---\\
209 		& 35, 105  & 207---209 (209)	& 28\\
209---211	& 40, 340  & 208---212 (211)	& 204\\
211---212 & 45, 170 & not detected &---\\
212---213 & -30, 125 & 212		& 24\\
214		& -20, 285 & 215---216 (215) & 99\\
214+	& -20, 310 & 213---217 (217) & 63\\
214+	& 55, 15    & 210---215 (214) & 32\\
\tableline
False detections:& & 201---204 (201) & 17\\
& & 205---206 (205) & 10\\
& & 207---208 (207) & 10\\
& & 212 & 4\\
\enddata
\label{tab:testperiod}
\tablenotetext{a}{Approximate location of manually-detected ejection.}
\tablenotetext{b}{Day with most selected points is given in parentheses.}
\tablenotetext{c}{Total number of selected points in detected ejection cluster over all days.}
\end{deluxetable*}

\begin{figure*}
\includegraphics[width=\textwidth]{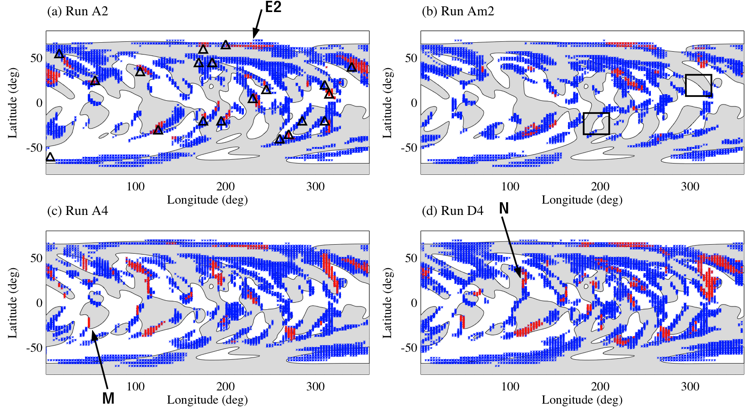}
\caption{Locations of flux rope ejections during the 13-day period from day 202 to day 214, used to test the automated detection algorithm. All flux rope points during the period are shown for runs (a) A2, (b) Am2, (c) A4, and (d) D4. Points are colored red if any point at that $(\theta,\phi)$ location was selected as part of an ejection by the algorithm, blue otherwise. The points are projected on to the normal photospheric magnetic field $B_{0r}$ from day 208 (white for positive and grey for negative). In panel (a), black triangles show manually-determined locations of ejections for run A2.}\label{fig:maxvr}
\end{figure*}

An immediate problem is how to define a flux rope ejection within our simulations. We adopt the practical definition of a large enough radial velocity in the magneto-frictional code. Thus we include both losses of equilibrium following gradual build-up of axial field, and sudden rises caused by nearby bipole emergence. We also include partial lift-offs where only one end of a flux rope opens up. Typically the other end is held down by overlying field from nearby regions. The majority of ejections remove the main, twisted, part of the flux rope through the top boundary of the computational box, usually within several days of the onset of rapid acceleration. 

Our automated procedure to find ejections is straightforward:
\begin{enumerate}
\item Record the radial velocity $v_{0r}$ in the magneto-frictional code at each of the selected flux rope points over the simulation, and select those points where $v_{0r}>0.5\,\textrm{km}\,\textrm{s}^{-1}$.
\item Cluster these points into separate ejection events.
\end{enumerate}
Although a similar algorithm is used, this clustering procedure differs from that used to define flux ropes on any particular day. Firstly, the clustering is carried out both in space {\it and time}, as ejections can take up to a few days to leave the computational domain. Secondly, for the distance measure we consider only the heliocentric angle between points, neglecting any radial distance. This allows for radial movement between daily snapshots. We require a minimum inter-group angle of $4\Delta$ between clusters, and a minimum separation in time of 4 days. After clustering, all groups of points with fewer than 4 points are discounted. These optimal parameter values have been determined by testing the detection procedure over a 13-day test period from day 202 to day 214 in run A2. Locations and days of detected ejections in this test period were compared to those found in a careful manual study of the magnetic field structure. The performance of the automated detection algorithm in this period (with the optimal choice of parameters found) is demonstrated by Table \ref{tab:testperiod}, and also illustrated in Figure \ref{fig:maxvr}(a), which shows all flux rope points during the period superimposed on the normal photospheric magnetic field for day 208. Panels (b), (c), and (d) correspond to other simulation runs and will be discussed in Section \ref{sec:ejections}. The first column of Table \ref{tab:testperiod} lists the days of significant vertical motion of the flux rope in each actual ejection, as detected manually. The approximate locations in latitude and longitude are listed in the second column, and plotted as black triangles in Figure \ref{fig:maxvr}(a). The third and fourth columns of Table \ref{tab:testperiod} give the results of the automated detection algorithm (with the optimal parameters). The clusters of points selected in this procedure are colored red in Figure \ref{fig:maxvr}(a). Note that only points falling within the 13-day period are shown in the figure, even if the cluster overlaps with the period. In Table \ref{tab:testperiod}, the number of clustered points refers to the total number in that cluster, whether in the 13-day period or not. The code detects 16 of the 19 ejections (84\%), and the number of clustered points (fourth column) gives an estimate of the size of each event. Of the three missed ejections, two are small, and the third (located at -60, 5) is quite large but rises slowly and gradually, with no sudden loss of equilibrium. We note that there are also four ``false positives,'' where spurious ejections are detected. However, two of these are small ``flux rope'' structures associated with temporary coronal reconfiguration following new bipole emergence, and the other two were actually considered on manual inspection to be part of other ejections. We conclude that this procedure performs well and the number of ejections detected is accurate to within $\pm 15\%$.

\section{Results} \label{sec:results}

\begin{figure*}
\includegraphics[width=\textwidth]{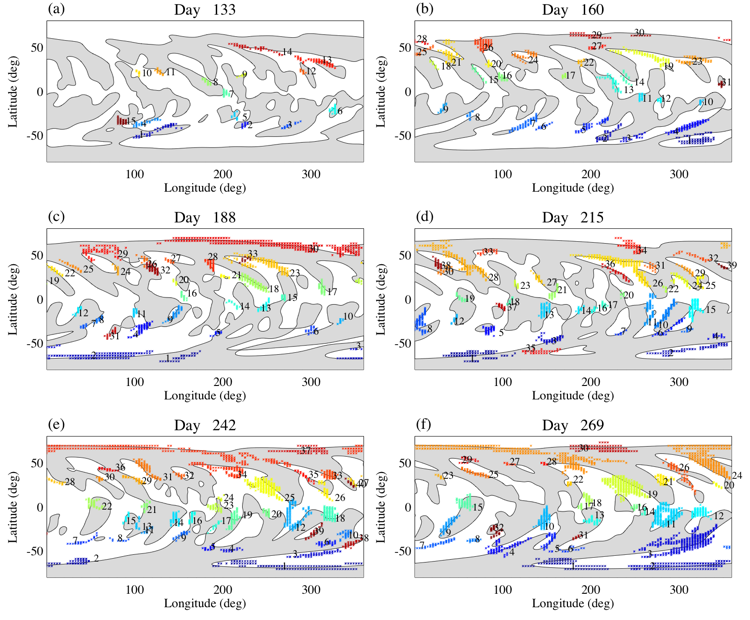}
\caption{Evolution of flux ropes during run A2, showing results of the automated procedure on six representative days. As in Figure \ref{fig:detection}(b), detected ropes are shown by colored points, projected on the normal magnetic field $B_{0r}$ in the photosphere (white shading for positive, grey for negative).}\label{fig:sequence}
\end{figure*}

We have applied the automated detection algorithms described in Section \ref{sec:auto} to each of the 177-day simulation runs in Table \ref{tab:runs}. In this section we discuss the results, concentrating first on overall properties of the flux ropes in each run (Section \ref{sec:param}), and then on the number of ejections in each run (Section \ref{sec:ejections}).

To give an overall feeling for the evolution, Figure \ref{fig:sequence} shows a sequence of snapshots of the flux ropes in simulation run A2 at 27-day intervals, as detected by the automated procedure in Section \ref{sec:autoropes}. Note that the number and size of flux ropes increase early in the simulation as the magnetic field diverges from the initial potential field, before the numbers begin to saturate. Their locations evolve as new bipoles emerge and the coronal field is reconfigured by surface motions.

\subsection{Dependence of Flux Rope Properties on Simulation Parameters} \label{sec:param}

\begin{figure}
\includegraphics[width=\columnwidth]{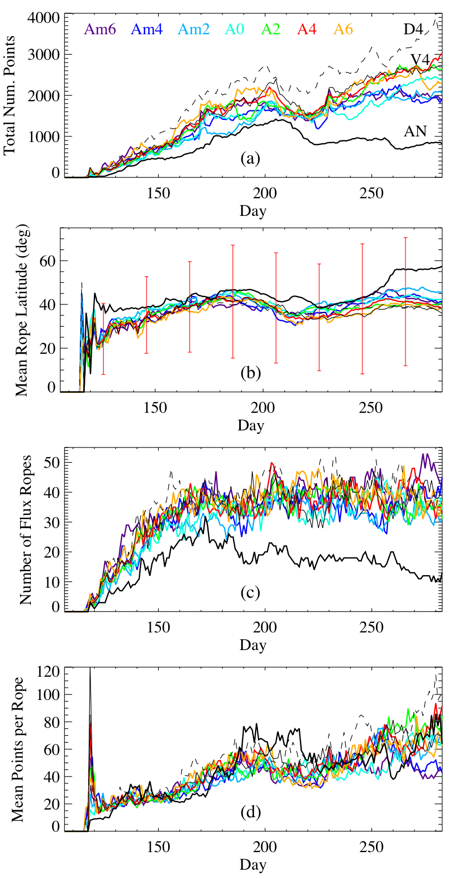}
\caption{Dependence of flux rope statistics on simulation parameters: (a) total number of points selected (on testing grid) after clustering, (b) mean latitude of selected points, (c) number of flux ropes present, and (d) mean number of points (on testing grid) per flux rope. Colored lines show runs with different emerging bipole twist $\beta$ (see legend in panel a). Thick black line shows run AN, thin solid black line shows run V4, and thin dashed black line shows run D4. In panel (b) vertical bars show $1\sigma$ for run A4.}\label{fig:statistics}
\end{figure}

\begin{figure*}
\includegraphics[width=\textwidth]{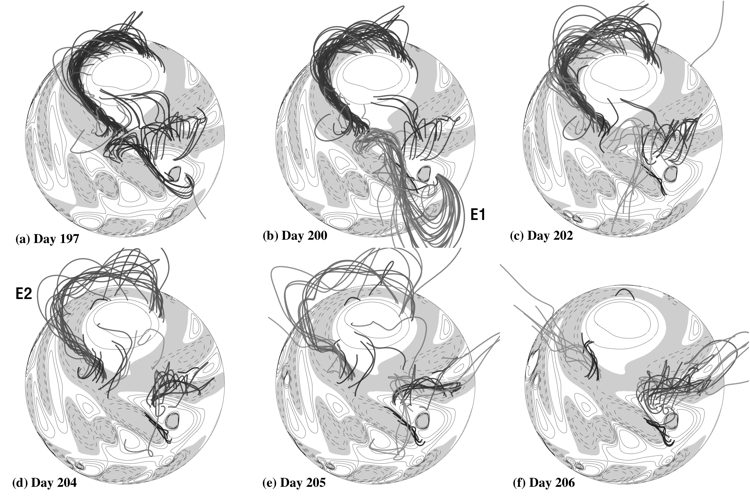}
\caption{Sequence of two high-latitude ejections (E1 and E2) in run A2, seen on days (a) 197, (b) 200, (c) 202, (d) 204, (e) 205, and (f) 206. Grey shading and thin contours show radial magnetic field $B_{0r}$ on photosphere (white/solid contours for positive, grey/dashed contours for negative). Thick lines show selected coronal magnetic field lines traced from flux rope points, with lighter grey denoting open field lines.}\label{fig:polar}
\end{figure*}

Figure \ref{fig:statistics} shows how the properties of flux ropes evolve over time in each simulation run. These properties include the total number of flux rope points (Figure \ref{fig:statistics}a), the mean latitude of flux rope points (Figure \ref{fig:statistics}b), the number of flux ropes found by the clustering stage (Figure \ref{fig:statistics}c), and the mean size of each of these clusters (Figure \ref{fig:statistics}d).

A general feature in all of these properties is an initial increase, followed in most cases by a saturation after 50 or more days. This illustrates the gradual build-up of helicity in the corona as the simulation evolves away from the initial potential field. Because run AN (with no bipole emergence) also shows an initial increase in the number of flux rope points (Figure \ref{fig:statistics}a), we see that surface shearing alone creates flux ropes over this timescale. However, the faster increase for the other runs shows that emerging helicity speeds up the formation of flux ropes. Indeed, Figure \ref{fig:statistics}(b) shows that more flux ropes form at lower latitudes in the runs with bipole emergence, supporting this point. The emergence of new flux also acts to sustain the coronal magnetic field against the diffusive decay which would otherwise occur over hundreds of days. This decay is evident in run AN as reformation of ejected flux ropes and production of new flux ropes declines after day 210.

Notice that all runs show a temporary decrease in the number of flux rope points and in their mean latitude around day 210. This is caused by the ejection of a pair of high-latitude flux ropes, illustrated in a time sequence in Figure \ref{fig:polar} (for run A2). The two ejections are labelled E1 and E2, and take place from days 199 and 204 respectively. Ejection E2 is also seen in Figure \ref{fig:maxvr}(a), but E1 occurred before the period shown there. A similar high-latitude ejection causes a dip in the number of flux rope points around day 260 for run AN.

\subsubsection{Effect of Emerging Bipole Twist}

Consider the effect of the emerging bipole twist $\beta$, as shown by colors in Figure \ref{fig:statistics}. Firstly, the number of flux ropes present at any time (Figure \ref{fig:statistics}c) saturates at around 38 in all runs, although each shows considerable fluctuation between about 28 and 48. There is some tendency toward more flux ropes in runs A6 and Am6, but this is not clear at all times. Run A0, where bipoles emerge untwisted, still has a comparable number of flux ropes, thus suggesting that the maintenance of photospheric flux is sufficient to produce flux ropes, without requiring the emergence of already-twisted fields.

There are, however, differences in the mean latitude and mean size of flux ropes in runs with different $\beta$. Over the full simulation, the mean latitude is lower for runs with more emerging bipole twist ($28.8^\circ$ for run A6 and $27.3^\circ$ for run Am6 {\it versus} $31.1^\circ$ for run A0). This suggests that greater emerging helicity accelerates flux rope formation at lower latitudes, although it doesn't significantly affect the number of locations at which flux ropes form.

Figure \ref{fig:statistics}(d) shows that, later in the simulation, the flux ropes are larger in the runs with majority hemispheric sign of bipole twist (A2, A4, and A6) as compared to the runs with opposite sign of twist (Am2, Am4, Am6). A more detailed analysis shows that the flux ropes that are larger in the first three runs are located mainly at mid-latitudes ($40^\circ$ to $60^\circ$), rather than at high latitudes or among newly-emerged active regions. The explanation is suggested by our analysis of sheared magnetic fields in filaments (Paper III). In particular, consider the PIL internal to a single bipole. Initially, the same amount of shear will emerge whether $\beta=-0.6$ or $\beta=0.6$, although with opposite sign. Hence there is no difference in flux rope sizes between runs Am6 and A6 at active latitudes. However, differential rotation always acts to produce the observed majority chirality (direction of shear) over such north-south oriented PILs. In run A6, the emerged shear already has the majority direction, but in run Am6 the emerged shear has the opposite direction, so the shear will first be reduced before the direction is reversed. Thus at mid-latitudes, {\it i.e.}, among older regions, the amount of shear and hence the size of the flux ropes will be lower in run Am6 than in run A6. There is no difference at high latitudes because, in the 177-day simulations, these are not influenced by the emerging regions.

In summary, the main effects of emerging bipole twist are to speed production of flux ropes at active latitudes for greater magnitude of $\beta$, and to produce larger flux ropes at mid-latitudes for the correct sign of $\beta$. The number of flux ropes present at any time is not affected in a systematic way.

\subsubsection{Effect of Radial Outflow}

In run V4, shown by the thin solid black line in Figure \ref{fig:statistics}, we halve the magnitude of the radial outflow speed to $50\,\textrm{km}\,\textrm{s}^{-1}$. We find no significant effect on the overall flux rope properties. This is because the outflow speed has no direct influence on the formation of flux ropes, or on the initial stages of the ejection. It acts only at the top of the computational domain, so influences only the (unphysical) evolution after equilibrium has been lost. To minimize dependence on this unphysical evolution after loss of equilibrium, our automated detection algorithm considers only flux ropes below $r=1.44R_\odot$.

\subsubsection{Effect of Coronal Diffusion}

The dashed black line in Figure \ref{fig:statistics} shows run D4, where the coronal diffusivity $\eta_0$ has only half of its usual value. This results in 20\% more flux rope points being found (Figure \ref{fig:statistics}a), and in 20\% larger flux ropes (\ref{fig:statistics}d). This is because the coronal diffusion acts to dissipate the concentrated currents in twisted flux rope structures. Lower diffusion allows more highly twisted structures to form. However, the coronal diffusion is not the primary source of axial magnetic flux above PILs, thus does not determine the locations at which flux ropes form. Figure \ref{fig:statistics}(c) shows that the mean latitude of flux ropes is not affected by the coronal diffusivity. A natural result of the larger, more twisted flux ropes formed in run D4 is to produce more ejections. We consider the number of ejections produced in different runs next.

\subsection{Rate of Flux Rope Ejections} \label{sec:ejections}

\begin{deluxetable}{cr}
\tablewidth{0pt}
\tablecaption{Rates of simulated ejections and observed CMEs.}
\tablehead{
\colhead{Simulation Run} & \colhead{Ejections per Day\tablenotemark{a}}}
\startdata
AN & $0.09\pm0.01$\\
Am6 & $1.05\pm0.16$\\
Am4 & $0.88\pm0.13$\\
Am2 & $0.61\pm0.09$\\
A0 & $0.67\pm0.10$\\
A2 & $1.01\pm0.15$\\
A4 & $1.20\pm0.18$\\
A6 & $1.28\pm0.19$\\
D4 & $1.60\pm0.24$\\
V4 & $1.16\pm0.17$\\
\tableline
LASCO events:&\\
All & 3.59\\
``Poor events'' removed & 2.25\\
\enddata
\label{tab:ejectionrate}
\tablenotetext{a}{Between day 183 (1999 July 2) and day 283 (1999 October 10).}
\end{deluxetable}

\begin{figure}
\includegraphics[width=\columnwidth]{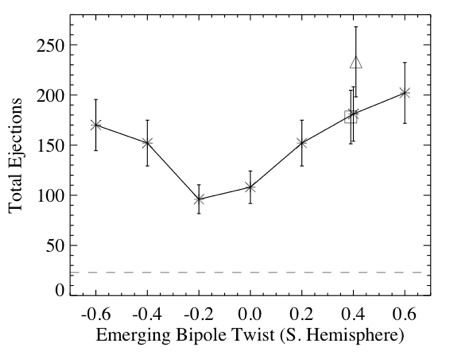}
\caption{Total number of ejections in different simulation runs, as detected by the automated procedure. Asterisks joined by solid line show runs Am6, Am4, Am2, A0, A2, A4, and A6. Dashed line shows run AN (where emerging bipole twist is irrelevant), triangle run D4, and square run V4. Error bars show estimated error in automated detection algorithm.}\label{fig:totalliftoffs}
\end{figure}

Using the automated procedure in Section \ref{sec:autoeject}, we find the number of ejections per day to have a ramp-up phase of about 40 days, before fluctuating considerably over the rest of the simulation, as well as varying between runs. Figure \ref{fig:totalliftoffs} shows the total number of ejections in each run, with error bars showing the estimated errors in the automated detection procedure (Section \ref{sec:autoeject}). We immediately see the importance of emerging bipoles for flux rope ejections. In run AN there are only $23\pm4$ ejections, as compared to $108\pm16$ in run A0, and more for the runs with emerging bipole twist---up to $202\pm30$ for run A6. Partly, this reflects the increased number of flux ropes in runs with emerging bipoles, and partly it reflects the influence of strong emerging magnetic fields in reconfiguring the coronal field. The increasing number of ejections with greater emerging bipole twist reflects the quicker formation of flux ropes with strong axial fields, which then lose equilibrium. For example, compare panels (a) and (c) in Figure \ref{fig:maxvr}, which show flux rope points and radial velocities for runs A2 and A4. The flux rope labelled M in Figure \ref{fig:maxvr}(c), internal to a newly-emerged bipole, both forms and is ejected in run A4. However, in run A2 there is not sufficient axial magnetic field at this location for a flux rope to be detected at this time.

 There is also a tendency for the runs with the majority hemispheric sign of bipole twist (A2, A4, and A6) to have more ejections than the runs with the same magnitude but opposite sign of twist. This may be related to the larger size of mid-latitude flux ropes in the first set of runs, as described in Section \ref{sec:param}. This is illustrated by comparing run Am2 with run A2 in Figure \ref{fig:maxvr} (panels b and a respectively). The black boxes in Figure \ref{fig:maxvr}(b) show decaying bipolar regions where no flux ropes are detected in run Am2. In run A2, both of these locations have already formed strong flux ropes which are ejected during this particular correlation period.

From run V4---shown by a square in Figure \ref{fig:totalliftoffs}---we see that halving the outflow speed has no significant influence on the number of ejections (giving $178\pm27$). So, just as the outflow has no major influence on the formation of flux ropes, it has no major influence on their loss of equilibrium. As before, this is because it acts only near the upper boundary, which a flux rope reaches only after equilibrium is lost. In contrast, halving the coronal diffusivity $\eta_0$ increases the number of ejections to $233\pm35$ in run D4 (triangle in Figure \ref{fig:totalliftoffs}). An example of an additional ejection not occurring in run A4 is labelled N in Figure \ref{fig:maxvr}(d). The higher number of ejections is explained by the larger, more twisted flux ropes which are able to form in this case.

Finally, how does the number of ejections produced in our simulations compare with the number of CMEs observed on the real Sun during this period? The CDAW catalog \citep{yashiro2004} is the standard manually-compiled list of CMEs observed by the SOHO/LASCO coronagraphs \citep{brueckner1995}. To avoid the initial ramp-up phase in the simulation, we compare the rates of ejections between day 183 (1999 July 2) and day 283 (1999 October 10). Table \ref{tab:ejectionrate} shows the number of flux rope ejections per day for each simulation run over this period, in addition to the observed rate from the CDAW catalog. Two observed rates are given: the first includes all events in the catalog, and the second omits events labelled ``poor'' by the LASCO operator. We believe the second rate to be more appropriate for this comparison, as our global simulations relate to large-scale flux rope events, which are unlikely to be labelled ``poor''. This rate will be a lower estimate for two reasons: firstly not all far-side events are observed, and, secondly, there are several gaps in the instrument coverage over the period. Using this rough estimate, our simulations produce flux rope ejections at about 50\% of the observed CME rate (after the initial ramp-up phase).

\section{Conclusion} \label{sec:conclusions}

We have studied the formation and ejection of magnetic flux ropes in a simplified model of the coronal magnetic field evolution, to begin to address the question of where and when CMEs occur in the global context. Loss of equilibrium of magnetic flux ropes in the low corona is a leading model for the initiation of CMEs, and the model described in this paper is, in essence, an extension of previous 2D catastrophe models of a single eruption to the 3D global corona. Ultimately, we aim to determine what proportion of observed CMEs may be explained by the loss of equilibrium mechanism. Using automated detection algorithms, we have tracked the formation, loss of equilibrium, and ejection of flux rope structures forming at many locations in the simulated corona, which evolves continually in response to the emergence of twisted bipolar active regions, and to large-scale motions on the photospheric boundary.

In this paper we consider the effect of key simulation parameters on the flux ropes formed and on the rate of flux rope ejections. We draw the following main conclusions:
\begin{enumerate}
\item The number of flux ropes present at any one time fluctuates between about 28 and 48, with no systematic dependence on the helicity of emerging bipoles. If no new bipoles are emerged, the surface flux decays and the number of flux ropes decreases due to ejections.
\item The magnitude of emerging bipole helicity has no major effect on the number of flux ropes present at any one time, but greater emerging helicity leads to more flux rope ejections.
\item The sign of emerging helicity also has an effect. If active regions emerge with the (observed) minority sign of twist in each hemisphere, then smaller flux ropes are created at mid-latitudes, and there are fewer ejections. This is because the surface shearing has first to reverse the direction of the sheared field that emerged with the minority sign of helicity, before forming a flux rope with the majority helicity.
\item The results are not sensitive to the upper boundary condition (radial outflow), but do depend on the turbulent diffusivity in the corona. A lower diffusivity leads to larger flux ropes and more ejections.
\item The rate of flux rope ejections in the model is roughly 50\% of the observed LASCO CME rate, depending on the choice of parameters in the model.
\end{enumerate}

We have shown that the rate of flux rope ejections varies from $0.67\pm0.10$ per day in run A0 (when bipoles emerge untwisted) to $1.28\pm0.19$ in run A6 (the greatest amount of emerging helicity considered). Since we do not, at present, have reliable measurements of the magnetic helicity in all 119 active regions that emerged during the simulated period, we cannot predict this ejection rate precisely. However, constraints from filament chirality observations (Paper II) suggest that simulation runs A4 and A6 are likely to be most realistic. Hence our conclusion that the quasi-static model can produce about 50\% of observed CMEs. Since the sign and magnitude of helicity emerging in active regions play an important role, more realistic future models must ultimately incorporate observations of helicity in individual emerging regions, which may come from the forthcoming Solar Dynamics Observatory (SDO) mission. Note that the rate of ejections in the model may be increased somewhat by reducing the turbulent diffusivity in the corona, although too small a value (as in run D4) produces flux ropes which are more highly twisted than typically observed. Future simulations will consider the effect of a higher-order form of  ``hyperdiffusion,'' which has been suggested to be more appropriate for the solar corona because it conserves magnetic helicity \citep{vanballegooijen2008}. 

Why does the model produce only 50\% of the observed CME rate? It is possible that the model over-estimates the amount of axial magnetic flux removed in ejections, and that in reality more axial flux is left behind, leading to another ejection from the same location shortly after. However, the axial flux rebuilds over a period of tens of days following an ejection, so this quasi-static model may never be able to produce a rapid succession of flares and eruptions from the same active region over minutes to hours, as is sometimed observed \citep{gopalswamy2006}. It seems likely, therefore, that up to 50\% of CMEs require either more complex magnetic field structures than allowed for in our simple model, or some physical mechanism other than loss of equilibrium of flux ropes formed quasi-statically by flux cancellation. One possibility is the dynamic emergence of already twisted flux ropes, and subsequent activity \citep[{\it e.g.},][]{tanaka1991}. Alternatively, in full MHD, eruptions may be triggered by a dynamical instability \citep{tokman2002,torok2003,kusano2004,kliem2006}.

Finally, in this paper we have concentrated on the ejection of flux ropes within the model, making only a rough comparison with the rate of observed CMEs. A future paper will describe a more detailed comparison with the source regions of observed CMEs, in order to determine whether the simplified global model can help to explain or predict the initiation locations of these events. In addition, it will discuss how meaningful present comparisons to observations are, and what extra input is required from observations so that more meaningful comparisons may be made. Only through such studies will we gain the insight of what is needed to predict CMEs in future.

\acknowledgments
We thank A.A. van Ballegooijen, P.C.H. Martens, and the anonymous referee for useful suggestions that have led to significant improvements in the paper's presentation. The simulations were performed on the UKMHD parallel computer in St Andrews, funded jointly by SRIF/STFC. ARY acknowledges financial support from NASA/LWS grant NNG05GK32G, from the UK STFC, and from the Solar Physics NSF-REU program at Montana State University, during which this work was initiated. DHM would like to thank the STFC for financial support, and the Royal Society for support through their research grant scheme. Synoptic magnetogram data from NSO/Kitt Peak were produced cooperatively by NSF/NOAO, NASA/GSFC, and NOAA/SEL and made publicly accessible on the World Wide Web.


\end{document}